\documentclass[a4paper]{article}

\usepackage{hyperref}
\usepackage{INTERSPEECH2022}
\usepackage{mathtools}

\title{Estimation of speaker age and height from speech signal using bi-encoder transformer mixture model}
\name{Tarun Gupta$^1$, Duc-Tuan Truong$^2$, Tran The Anh$^2$, Chng Eng Siong$^2$}
\address{
  $^1$Department of Computer Science and Engineering, Indian Institute of Technology, Indore, India\\
  $^2$School of Computer Science and Engineering, Nanyang Technological University, Singapore}
\email{cse1800001059@iiti.ac.in, \{DUCTUAN001, TRANTHEA001\}@e.ntu.edu.sg, ASESChng@ntu.edu.sg}

\begin{document}

\maketitle
\begin{abstract}
The estimation of speaker characteristics such as age and height is a challenging task, having numerous applications in voice forensic analysis. In this work, we propose a bi-encoder transformer mixture model for speaker age and height estimation. Considering the wide differences in male and female voice characteristics such as differences in formant and fundamental frequencies, we propose the use of two separate transformer encoders for the extraction of specific voice features in the male and female gender, using wav2vec 2.0 as a common-level feature extractor. This architecture reduces the interference effects during backpropagation and improves the generalizability of the model. We perform our experiments on the TIMIT dataset and significantly outperform the current state-of-the-art results on age estimation. Specifically, we achieve root mean squared error (RMSE) of 5.54 years and 6.49 years for male and female age estimation, respectively. Further experiment to evaluate the relative importance of different phonetic types for our task demonstrate that vowel sounds are the most distinguishing for age estimation.\footnote{Code and models are available at \href{https://github.com/tarun360/SpeakerProfiling}{\textit{github/tarun360/SpeakerProfiling}}}


\end{abstract}
\noindent\textbf{Index Terms}: speaker profiling, age estimation, height estimation, mixture of experts.

\section{Introduction}
Speech is an acoustic output produced by precisely coordinated movement of different human body parts. Therefore, there have been suggestions that acoustic features of speech can convey information about the physical characteristics of the speaker. Among various physical characteristics, scientific studies have investigated the correlation between voice characteristics and a speaker's age and height. Authors of \cite{Arsikere14-SHSR, singh2016short} reported that the vocal tract length, sub-glottal resonance frequencies, and formant frequencies are correlated with the individual’s height. Other voice characteristics of speech such as speech rate, sound pressure level, fundamental frequency, etc.\ vary according to the speaker’s age \cite{singh2016short,Schotz07-AAASPA}. The age-related glottis deterioration of the speaker also impacts the speech characteristics like jitter, shimmer \cite{Mller07-CSC}, and speech harmonics \cite{Li12-ASAGR}.

Automatic speaker profiling systems could be applied to a variety of different fields. For example, in criminal investigation, evidence could be in the form of voice recordings, e.g. a hoax bomb threat or ransom demand over a phone call \cite{PHC-2016, Schuller2013ParalinguisticsIS}. In such cases, estimating the age and height of speakers in the audio evidence could save investigating agencies' time by narrowing down the number of suspects. Similarly, predicting the age and gender of a speaker from the speech data can aid marketing campaigns by targeting suitable gender/age customer groups \cite{Schuller2013ParalinguisticsIS}. In addition, the speaker profiling system can benefit other speech fields, namely, speaker diarization and speech-based verification as well.



\textbf{Contribution:} In this work, we study the use of self-supervised speech representation, specifically wav2vec 2.0 \cite{baevski2020wav2vec}, for speaker age and height estimation. We also compare wav2vec 2.0 to other famous feature representations: MFCC, filter bank, and x-vectors \cite{snyder2018x}. To our best knowledge, our work is the first in the literature to demonstrate the potential of wav2vec 2.0 in predicting speaker age and height. For the downstream architecture, we present a novel Mixture of Experts (MoE) based bi-encoder transformer model that utilizes wav2vec 2.0 as a common-level speech feature extractor followed by a bi-encoder architecture. The bi-encoder architecture is motivated from the differences in male and female voice characteristics such as differences in formant and fundamental frequencies \cite{GRP-1991, MENDOZA1996}. The two encoders act as `experts' for providing distinctive features for male and female voice. We further make use of homoscedastic uncertainty \cite{kendall2018multi} to define the multi-task loss function and mixup \cite{mixup2017} as a regularization technique. The proposed network achieve new state-of-the-art results in age estimation on the TIMIT dataset.




The rest of our paper is organized as follows. Section 2 briefly describes the speaker profiling literature. Section 3 illustrates our proposed model architecture and techniques. Section 4 describes the experiments conducted to estimate the age and height of a speaker in a monolingual setting using the TIMIT dataset. Also, this section discusses the impact of the different phones on the estimation result in Section 4.3. Finally, conclusions are presented in Section 5.

\begin{figure*}[t]
  \centering
  \includegraphics[width=0.65\linewidth]{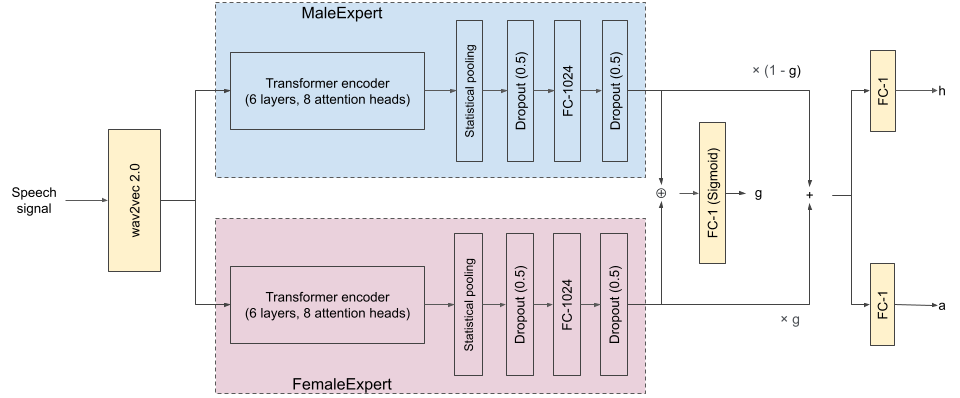}
  \caption{wav2vec 2.0 bi-encoder model architecture. Transformer encoder \cite{vaswani2017attention} with 6 layers and 8 attention heads is used. Statistical pooling refers to taking mean and standard deviation along frame dimension and concatenating them to obtain utterance level representation. `FC-L' denotes a fully connected layer with L neurons. $\oplus$ denotes the concatenation operation. a, h and g denote the age, height and gender prediction.}
  \label{fig:model_arch}
\end{figure*}

\section{Related works}
There is a wide choice of speech feature extractors for automatic height and age estimation available in the literature. However, most of the previous studies used conventional techniques of extracting features from raw speech signals. In the earlier years, authors of \cite{mporas2009estimation, AFS-2010} used the Open-Smile toolkit to convert the short-term spectral features into various statistics like mean, median, percentiles, etc.\ for height and age estimation. A similar statistical approach for age and height prediction applied i-vector (dimension reduced version of Gaussian mixture model universal background model) \cite{SAE-16, AES-18} to convert a variable-length utterance into a fixed-size embedding vector. Another embedding approach of speaker profiling is obtaining spectral characterizations of the speech signal. For example, the speaker's resonance frequencies of sub-glottal are used for height estimation \cite{Arsikere2013AutomaticEO}. Singh et al. \cite{singh2016short} applied a bag of words representation to capture short-term cepstral features with different time resolutions. Other common-level short term features are Mel Frequency Cepstral Coefficients (MFCC) \cite{VAAC-97, kalluri2019deep}, cepstral and pitch features \cite{Mller07-CSC, ESH-21}, etc. 

 In this era of deep learning, deep neural networks (DNN) have shown the outstanding ability to discover descriptive and distinctive representations from raw speech audio. Therefore, recent studies have applied DNN-based speech representation learning to improve the accuracy of the speaker's height and age prediction. Abumallouh et al.~\cite{NTF-2018} presented a speaker age and gender classifier that based on top of an unsupervised DNN bottleneck feature extractor can achieve better performance than its original MFCCs feature set, especially for female speaker. By applying DNN discriminative embedding called x-vector \cite{xvector-2018}, authors of \cite{kwasny2020joint, GAE-2021} gained better results in height and age estimation in TIMIT dataset compared to previous studies. Shangeth et al.~\cite{rajaa2021learning} employed semi-supervised learning approach to learn speaker representation and achieve the state-of-the-art result on age estimation on the TIMIT Test dataset with Mean Absolute Error (MAE) of 4.8 and 5.0 years for male and female speakers respectively. Recently, Baevski et al.~\cite{baevski2020wav2vec} presented the wav2vec 2.0 framework for self-supervised learning of discrete speech units. In the field of speaker recognition, wav2vec 2.0 have increased the classification accuracy considerably since it can capture much more phonetics information than it's conventional counterpart. To the best of our knowledge, there is no work in the literature which studies the use of wav2vec 2.0 for speaker height and age estimation. 
 
Literature also showed that there are differences between male and female voices. The average female formant and fundamental frequencies are higher than that of the male speaking voice \cite{GRP-1991, MENDOZA1996}. Therefore, the extraction of height and age information in the speech signal is also affected by the gender of the speaker \cite{KALLURI2020}. Most of the previous speaker profiling studies considered gender classification and height/age estimation as a joint problem \cite{GAE-2021, rajaa2021learning}. The only work to use gender information and get a slight improvement in speaker's height and age prediction is by \cite{ESH-21, kaushik2021end}. Nevertheless, these works adapted ground truth gender information which is fed to their model just as a binary value. In order to capture distinctive representations of male and female speech, we use two experts, one for each gender. Mixture of Experts (MoE) \cite{jacobs1991adaptive} is a supervised learning system in which separate networks are designed to handle different subsets of the training samples. Recently, the MoE architecture has been studied for various tasks in the speech domain such as multi-accent speech recognition \cite{jain2019multi}, code-switching speech recognition \cite{lu2020bi}, etc. 

\section{Method}

\begin{table*}[t]
  \caption{Comparison of the proposed wav2vec 2.0 bi-encoder model with the existing work.}
  \label{tab:main}
  \centering
  \begin{tabular}{ c | c | c | c | c | c | c | c | c }
    \toprule
    \textbf{Method} & \multicolumn{2}{c}{\textbf{Height RMSE}} & \multicolumn{2}{c}{\textbf{Height MAE}} & \multicolumn{2}{c}{\textbf{Age RMSE}} &  \multicolumn{2}{c}{\textbf{Age MAE}} \\
     & Male & Female & Male & Female & Male & Female & Male & Female \\ 
    \midrule
    Singh et al. \cite{singh2016short} & \textbf{6.7} & \textbf{6.1} & \textbf{5.0} & \textbf{5.0} & 7.8 & 8.9 & 5.5 & 6.5 \\
    Kalluri et al. \cite{kalluri2019deep} & 6.85 & 6.29 & - & - & 7.60 & 8.63 & - & - \\
    Kwasny et al. \cite{kwasny2020joint} & - & - & - & - & 7.24 & 8.12 & 5.12 & 5.29 \\
    Williams et al. \cite{williams2013speaker} & - & - & 5.37 & 5.49 & - & - & - & - \\
    Mporas et al. \cite{mporas2009estimation} & 6.8 & 6.3 & 5.3 & 5.1 & - & - & - & - \\
    Shangeth et al. (single-task model) \cite{rajaa2021learning} & 8.1 & 6.0 & 5.9 & 4.9 & 6.96 & 7.6 & 4.8 & 5.1 \\
    Shangeth et al. (multi-task model) \cite{rajaa2021learning} & 7.5 & 6.5 & 5.8 & 5.1 & 6.8 & 7.4 & 4.8 & 5.0 \\
    Manav et al. (single-task model) \cite{kaushik2021end} & 6.92 & 6.24 & 5.20 & 4.95 & 7.20 & 7.10 & 5.04 & 5.02 \\ 
    Manav et al. (multi-task model) \cite{kaushik2021end} & 6.95 & 6.44 & 5.26 & 5.15 & 7.81 & 8.60 & 5.50 & 5.89 \\ 
    \textbf{wav2vec 2.0 bi-encoder (ours)} & 7.3 & 6.43 & 5.58 & 5.07 & \textbf{5.54} & \textbf{6.49} & \textbf{3.96} & \textbf{4.48} \\
    \bottomrule
  \end{tabular}
\end{table*}

\begin{table*}[t]
  \caption{Comparison of wav2vec 2.0 bi-encoder model with wav2vec 2.0 single-encoder model.}
  \label{tab:biEncoderComparison}
  \centering
  \begin{tabular}{ c | c | c | c | c | c | c | c | c }
    \toprule
    \textbf{Method} & \multicolumn{2}{c}{\textbf{Height RMSE}} & \multicolumn{2}{c}{\textbf{Height MAE}} & \multicolumn{2}{c}{\textbf{Age RMSE}} &  \multicolumn{2}{c}{\textbf{Age MAE}} \\
     & Male & Female & Male & Female & Male & Female & Male & Female \\ 
    \midrule
    wav2vec 2.0 single-encoder & \textbf{7.17} & \textbf{6.39} & \textbf{5.35} & 5.08 & 5.71 & 6.98 & 4.05 & 4.90 \\
    \textbf{wav2vec 2.0 bi-encoder} & 7.3 & 6.43 & 5.58 & \textbf{5.07} & \textbf{5.54} & \textbf{6.49} & \textbf{3.96} & \textbf{4.48} \\
    \bottomrule
  \end{tabular}
\end{table*}

\begin{table*}[t]
  \caption{Comparison of different feature extractors: MFCC, filter bank, x-vectors and wav2vec 2.0.}
  \label{tab:commonFeatureComparison}
  \centering
  \begin{tabular}{ c | c | c | c | c | c | c | c | c }
    \toprule
    \textbf{Method} & \multicolumn{2}{c}{\textbf{Height RMSE}} & \multicolumn{2}{c}{\textbf{Height MAE}} & \multicolumn{2}{c}{\textbf{Age RMSE}} &  \multicolumn{2}{c}{\textbf{Age MAE}} \\
     & Male & Female & Male & Female & Male & Female & Male & Female \\ 
    \midrule
    MFCC bi-encoder & 7.63 & 6.69 & 5.79 & 5.33 & 8.15 & 8.65 & 5.86 & 6.02 \\
    filter bank bi-encoder & 7.86	& 6.68 & 6.13 & 5.36 & 8.51 & 8.42 & 6.19 & 5.86  \\
    x-vectors bi-encoder & 8.02	& 6.79 & 6.11 & 5.46 & 7.66 & 8.89 & 5.5 & 5.82  \\
    \textbf{wav2vec 2.0 bi-encoder} & \textbf{7.3} & \textbf{6.43} & \textbf{5.58} & \textbf{5.07} & \textbf{5.54} & \textbf{6.49} & \textbf{3.96} & \textbf{4.48} \\
    \bottomrule
  \end{tabular}
\end{table*}

\subsection{Self supervised representation}
Motivated by the success of self-supervised learning in the field of speech recognition, we explore an SSL model, namely wav2vec 2.0 \cite{baevski2020wav2vec} in speaker profiling. The wav2vec 2.0 model learns speech representations by solving contrastive tasks in latent space. Specifically, wav2vec 2.0 is composed of 1D convolution features extractor $f : X \rightarrow Z$ which expects raw audio input $X$ and outputs latent representations $Z$, followed by transformer encoders $e : Z \rightarrow C$, which provides contextual information. The feature extractor outputs are quantised by a quantisation module.

\subsection{Mixture of Experts}
The Mixture of Experts (MoE) \cite{jacobs1991adaptive} concept is based on divide-and-conquer principle. If the training dataset is known in advance to be divided naturally into certain subspaces, then separated experts of these subspaces can be trained. A gating network decides the weights to be assigned to each of the expert views and then a weighted sum of all the expert views is taken. The MoE architecture leads to less interference during backpropagation and can provide faster training and better generalizability. 

\subsection{Model architecture}

The model architecture has been described in Fig.\ref{fig:model_arch}. First, wav2vec 2.0 is used to extract features from the raw audio waveform. Then, utilising the MoE concept, a bi-encoder transformer network is built on top of the extracted features. For the two genders present in the TIMIT dataset, male and female, we consider separate experts, $MaleExpert$ and $FemaleExpert$. These two experts have identical architectures as illustrated in Fig. \ref{fig:model_arch}, having 6 layers of self-attention based transformer encoder with 8 attention heads in each layer \cite{vaswani2017attention}. We take an average and standard deviation of the output of the transformer encoder along the frame dimension and concatenate them to get utterance level representations. These utterance level representations are then fed to dropout and fully connected layers to complete the expert architecture. The two separate experts can focus on gender-based audio characteristics important for estimating height and age, while the fine-tuned wav2vec 2.0 can learn common-level features for both genders. 


The two experts, $MaleExpert$ and $FemaleExpert$ provide two expert views $e_m$ and $e_f$:
\begin{equation}
  e_m = MaleExpert(x)
\end{equation}
\begin{equation}
  e_f = FemaleExpert(x)
\end{equation}
where $x$ is the feature representation extracted from fine tuned wav2vec 2.0. $e_f$ and $e_m$ are concatenated and fed to a fully connected layer with sigmoid activation to predict the gender $g \in [0,1]$. Male gender is encoded as 0 and female as 1. Gender prediction acts as a gating network to combine the output of the two experts as follows:
\begin{equation}
  e = (1-g) \times e_m + g \times e_f
\end{equation}
$e$ is then fed to fully connected layers to perform age and height regression.

\subsection{Loss function}
In our multi-task model, three losses are factored into the training process, namely height, age and gender. Prior approaches for building a multi-task model for speaker profiling have used a weighted sum of these losses \cite{rajaa2021learning, kaushik2021end}, where the loss weights are manually fine-tuned. Instead, we use uncertainty loss \cite{kendall2018multi}, which uses homoscedastic uncertainty to combine multiple losses. Using this, we define our multi-task loss function as:
\begin{equation}
  \mathcal{L} = \frac{\mathcal{L}_{height}}{2\sigma^2_{height}} + \frac{\mathcal{L}_{age}}{2\sigma^2_{age}} + \frac{\mathcal{L}_{gender}}{2\sigma^2_{gender}} +  log(\sigma_{height}\sigma_{age}\sigma_{gender})
  \label{eq:lossfn}
\end{equation}
where $\sigma_{height}$, $\sigma_{age}$ and $\sigma_{gender}$ are learnable parameters. As suggested in their paper, we make the substitution $s_{height} = log(\sigma^2_{height})$, $s_{age} = log(\sigma^2_{age})$ and $s_{gender} = log(\sigma^2_{gender})$ in equation \ref{eq:lossfn} for numerical stability.

\subsection{Mixup}
We use mixup as a regularization technique to encourage the model to learn linear inference from the input audio. Mixup augmentation has been used previously in text-independent speaker verification \cite{zhu2019mixup}. 

Given two audio samples $x_i, x_j$, with $h_i, h_j$, $a_i, a_j$ and $g_i, g_j$ as the corresponding height, age, and gender values, the mixup augmented sample is defined as:
\begin{equation}
  x_{mixup} = \lambda x_i + (1-\lambda) x_j
\end{equation}
\begin{equation}
  h_{mixup} = \lambda h_i + (1-\lambda) h_j
\end{equation}
\begin{equation}
  a_{mixup} = \lambda a_i + (1-\lambda) a_j
\end{equation}
\begin{equation}
  g_{mixup} = \lambda g_i + (1-\lambda) g_j
\end{equation}
where $\lambda \sim U(0,1)$. To deal with audios of different lengths, we repeat the audio of a shorter length to make it of the same length as the longer audio. 
\begin{table}[t]
  \caption{Percentage change in RMSE values due to phoneme masking}
  \label{tab:phone_analysis}
  \centering
  \begin{tabular}{ c | c | c | c | c  }
    \toprule
    \textbf{Mask} & \multicolumn{2}{c}{\textbf{Height RMSE}} & \multicolumn{2}{c}{\textbf{Age RMSE}}  \\
     & Male & Female & Male & Female  \\ 
    \midrule
    Vowels & 2.04\% & 0.07\% & 38.9\% & 20.46\% \\
    Nasals & -0.52\% & 0.08\% & 2.51\% & -0.27\% \\
    Semivowels & 0.45\% & -0.32\% & 12.2\% & -0.68\% \\
    Affricates & 0.0\% & 0.0\% & 0.0\% & 0.0\% \\
    Fricatives & 1.28\% & 2.87\% & 6.08\% & -2.41\% \\
    Stops & 0.6\% & 2.9\% & 4.14\% & 3.05\% \\
    Others & 1.07\% & -2.9\% & 5.84\% & 12.17\% \\
    \bottomrule
  \end{tabular}
  
\end{table}

\section{Experiments}

\subsection{Dataset}
In this work, we use TIMIT dataset \cite{garofolo1993darpa} for our experiments. It consists of audio samples from 630 speakers, having eight different dialects of American English. The height varies in the range of 145cm to 204cm, and age varies in the range of 21 years to 76 years. TIMIT dataset comes pre-divided into train and test sets and we randomly select 15\% of train set for validation.

\subsection{Experimental Setup}
For the proposed bi-encoder transformer model, apart from wav2vec 2.0, we also consider other features for comparison: MFCC, filter bank and x-vectors. For filter bank, we consider 80 mel bins along with first and second order delta features. For MFCC, we consider 16 cepstral coefficients, along with first and second order delta features. For both filter bank and MFCC, we use apply Cepstral Mean and Variance Normalization (CMVN) and use frame length of 25ms and frame shift of 10ms. For x-vectors, we extract the frame-level features before it's statistical pooling layer, which is then fed to downstream bi-encoder architecture. The x-vectors were pre-trained on VoxCeleb1 \cite{nagrani2017voxceleb} and VoxCeleb2 \cite{chung2018voxceleb2} training data.

In another experiment, to show the efficacy of bi-encoder architecture, we compare the wav2vec 2.0 bi-encoder model with wav2vec 2.0 single-encoder model. The wav2vec 2.0 single-encoder model has the same layers as the wav2vec 2.0 bi-encoder model, except for the fact that there's only a single expert for both genders and there's no gating network. The output of the single expert is fed to fully connected layers to perform age, height, and gender estimation.

For models using wav2vec 2.0, we unfreeze the entire wav2vec 2.0 architecture except for the first five convolution layers of the convolutional feature extractor and use Adam optimizer \cite{kingma2014adam} with learning rate $10^{-6}$. When using MFCC, filter bank and x-vectors, we use Adam optimiser with learning rate $10^{-5}$. Mixup was proposed as regularization technique for deep neural network architectures to favour simple linear behaviour, and as such, cannot be applied directly at audio-level for traditional feature extraction techniques like MFCC and filter bank. As a result, we use mixup only when utilising x-vectors and wav2vec 2.0 as common feature extractor. 

\subsection{Experimental Results}
In Table \ref{tab:main}, we compare the results of the wav2vec 2.0 bi-encoder model with previous works. While many previous works built separate models for age and height, or separate models for male and female, we compare our multi-task model with all of them. We report our results both in root mean squared error (RMSE) and mean absolute error (MAE). It can be observed that the model achieves significant improvement in the age estimation task. Specifically, we achieve RMSE error 5.54 years and 6.49 years, which corresponds to relative improvement of 18.5\% in male age estimation and 8.6\% in female age estimation over the current state-of-the-art. 

The comparison of wav2vec 2.0 bi-encoder vs wav2vec 2.0 single encoder model has been illustrated in Table \ref{tab:biEncoderComparison}. The wav2vec 2.0 bi-encoder model achieves a relative improvement of 2.9\% for male age estimation and 7.0\% for female age estimation over wav2vec 2.0 single-encoder model in terms of RMSE error, testifying our hypothesis of using separate encoders for the two genders. However, the same pattern is not observed in height estimation results, indicating that the bi-encoder architecture is not as useful for height estimation as it is for age estimation.

In Table \ref{tab:commonFeatureComparison} we tabulate the results of different feature extractors: MFCC, filter bank, x-vectors and wav2vec 2.0. It can be observed that wav2vec 2.0 feature representation is superior to other feature extraction techniques.

\subsection{Phonetic importance analysis}
In order to understand the relative importance of the different types of phones in speaker profiling, we analyse the results of wav2vec 2.0 bi-encoder model after masking all utterances of a particular phone type. TIMIT dataset provides time-aligned phonetic transcriptions for each audio sample. Phones used in TIMIT corpus are divided into the following classes: `Stops', `Affricates', `Fricatives', `Nasals', `Semivowels and Glides', `Vowels' and `Others'. For each of these phone types, we mask all the phones in audio samples in TIMIT's test set, and then calculate the height and age RMSE scores. The relative percentage change due to these phone masking, as compared to when no masking was done, are tabulated in Table \ref{tab:phone_analysis}. We notice the largest increase in age RMSE value due to `Vowel' masking, implying vowel sounds are the most important phoneme type for age estimation. Surprisingly, we notice no significant change in height RMSE values, indicating that height estimation is not dependent on any particular phoneme type. 

\section{Conclusions}
In this work, we presented a Mixture of Experts (MoE) based bi-encoder transformer model, that uses self-supervised representation, specifically wav2vec 2.0, for speaker age and height estimation. The experimental results demonstrate that having separate experts for male and female voices can reduce interference during training process and can achieve state-of-the-art results for age estimation. We employed the homoscedastic uncertainty principle to combine multiple losses of our multi-task model. As part of future work, we plan to explore different feature extractors and self-supervised learning to further improve age and height estimation results.
\section{Acknowledgements}
The computational work for this article was partially performed on resources of the National Supercomputing Centre, Singapore (\href{https://www.nscc.sg}{https://www.nscc.sg}).

\bibliographystyle{IEEEtran}

\bibliography{template}

\end{document}